\documentclass{aa}
\usepackage{graphicx}

\usepackage{txfonts}

\newcommand{\f}{f^{(\nu)}}
\newcommand{\be}{\begin{equation}}
\newcommand{\ee}{\end{equation}}

\begin{document}
   \title{A family of models of partially relaxed stellar systems}

   \subtitle{ I. Dynamical properties}

   \author{M. Trenti\inst{1}
          \and
           G. Bertin\inst{2}
          }

   \offprints{m.trenti@sns.it}

     \institute{Scuola Normale Superiore, Piazza dei Cavalieri 7,
              I-56126 Pisa, Italy\\
              \email{m.trenti@sns.it}
         \and
             Dipartimento di Fisica, Universit\`{a} di Milano, via
Celoria 16, I-20133 Milano, Italy \\
             \email{Giuseppe.Bertin@unimi.it}
             }

   \date{Received, 2004 ; accepted}

   \abstract{Recently we have found that a family of models of
partially relaxed, anisotropic stellar systems, inspired earlier by
studies of incomplete violent relaxation, exhibits some interesting
thermodynamic properties. Here we present a systematic investigation
of its dynamical characteristics, in order to establish the basis for
a detailed comparison with simulations of collisionless collapse,
planned for a separate paper. For a full comparison with the
observations of elliptical galaxies, the models should be extended to
allow for the presence a sizable dark halo and of significant
rotation. In the spherical limit, the family is characterized by two
dimensionless parameters, i.e. $\Psi$, measuring the depth of the
galaxy potential, and $\nu$, defining the form of a third global
quantity $Q$, which is argued to be approximately conserved during
collisionless collapse (in addition to the total energy and the total
number of stars).

The family of models is found to have the following properties.
The intrinsic density profile beyond the half-mass radius $r_M$ is
basically universal and independent of $\Psi$. The projected
density profiles are well fitted by the $R^{1/n}$ law, with $n$
ranging from $2.5$ to $8.5$, dependent on $\Psi$, with $n$ close
to 4 for concentrated models. All models exhibit radial anisotropy
in the pressure tensor, especially in their outer parts, already
significant at $r \approx r_M$. At fixed values of $\nu$, models
with lower $\Psi$ are more anisotropic; at fixed values of $\Psi$,
models with lower $\nu$ are more concentrated and more
anisotropic. When the global amount of anisotropy, measured by
$2K_r/K_T$, is large, the models are unstable with respect to the
radial-orbit instability; still, a wide region of parameter space
(i.e., sufficiently high values of $\Psi$, for $\nu > 3/8$) is
covered by models that are dynamically stable; for these, the line
profiles (line-of-sight velocity distribution) are Gaussian at the
$5\%$ level, with a general trend of positive values of $h_4$ at
radii larger than the effective radius $R_e$.

   \keywords{stellar dynamics --- galaxies: evolution --- galaxies:
formation --- galaxies: kinematics and dynamics --- galaxies:
structure}
   }
   \authorrunning{M. Trenti \& G. Bertin}
   \maketitle
%

\section{Introduction}

A simple picture for the incomplete violent relaxation of stellar
systems considers the collapse of a dynamically cold cloud of
stars or star-clumps initially far from equilibrium. The study of
this ideal and relatively simple process, pioneered by important
analysis in the 60s (starting with \cite{lyn67}) and simulations
in the 80s (see \cite{van82}), is still incomplete. A proper
understanding of such process is a prerequisite to more ambitious
attempts at constructing physically justified models of elliptical
galaxies in which the problem of galaxy formation is set in the
generally accepted cosmological context of hierarchical
clustering.

Before the development of such cosmological scenarios, a first
step (in the direction of incorporating in a simple analytical and
physically justified framework the clues gathered from simulations
of collisionless collapse) had been taken in terms of the
so-called $f_{\infty}$ models (\cite{ber84}). These
models are constructed from a distribution function which, in the
spherical limit, reduces to $f_{\infty} = A(-E)^{3/2}\exp{(-aE -
cJ^2/2)}$, for negative values of $E$, and vanishes for positive
values of $E$; here $A$, $a$, and $c$ are positive constants and
$E$ and $J$ denote the specific star energy and angular momentum.

From those earlier investigations, it immediately became clear
that the physical clues gathered from the picture of collisionless
collapse (i.e., that incomplete violent relaxation leads to
systems that are well relaxed in their inner regions, $r \ll r_M$,
and characterized by radially anisotropic envelopes for $r \gg
r_M$) do not lead uniquely to the $f_{\infty}$ models, but instead
identify a wide class of attractive distribution functions of
which $f_{\infty}$ represents just one simple and interesting
case. At that time, the primary goal of a series of investigations
(described, e.g., by \cite{ber93}) was to test whether
the models inspired by studies of collisionless collapse were
realistic and thus could serve as a useful tool to interpret the
observations. Indeed, the $f_{\infty}$ models turned out to
``explain" the $R^{1/4}$ luminosity law (\cite{dev48}) and,
extended to the two-component case, were used successfully to
probe the presence and size of dark halos in elliptical galaxies.

Because of the focus on such astronomical applications, the
problem of a detailed comparison between the $f_{\infty}$ or other
models and the products of collisionless collapse from N-body
simulations was given lower priority and basically left aside.
Yet, it was pointed out that, with respect to the products of
numerical simulations of collisionless collapse, the $f_{\infty}$
models had the undesired feature of being too isotropic. Some
authors (\cite{mer89}) thus suggested that the $f_{\infty}$ models
should be extended to and used in the parameter domain where $a <
0$; however, this attempt failed, not only because a proper
physical justification was lacking, but especially because such
``negative-temperature" models suffer from the opposite
difficulty, i.e. they were shown to be so anisotropic that they
are violently unstable and bound to evolve on a time scale even
smaller than the typical crossing time.

In a recent paper (\cite{ber03}) we have revisited the problem of
the structure and dynamics of partially relaxed stellar systems
starting from a thermodynamic description. In view of the paradigm
of the {\it gravothermal catastrophe} (see \cite{ant62};
\cite{lyn68}; \cite{kat78,kat79,kat80}), we found it appropriate
to approach the problem in terms of the so-called $f^{(\nu)}$
models (for a definition, see following Sect.~2); with respect to
the $f_{\infty}$ models, the $f^{(\nu)}$ models follow from a
statistical mechanical derivation (\cite{sti87}) that is formally
straightforward, while they are also known to have generally
similar and reasonably realistic structural properties.  It should
be noted that the dynamical properties of the $f^{(\nu)}$ models
had not been studied earlier in detail. Such recent investigation,
focusing on the issue of the gravothermal catastrophe, convinced
us that the $f^{(\nu)}$ models do have many attractive features;
in particular, these models turn out to have a higher degree of
radial pressure anisotropy with respect to the $f_{\infty}$
models, making them more suitable to describe the results of
simulations of collisionless collapse. Such encouraging
preliminary inspection was the basis for a thorough study that we
have thus performed and that we present here.

In this paper we provide a systematic description of the dynamical
properties of the $f^{(\nu)}$ models. In the spherical limit,
these models define a {\it two-parameter family}, characterized by
the dimensionless parameter $\Psi$, measuring the depth of the
galaxy potential, and the dimensionless parameter $\nu$, defining
the form of a third global quantity $Q$ (which is argued to be
approximately conserved during collisionless collapse, in addition
to the total energy and the total number of stars; see Sect.~2 for
the relevant definitions). In Sect.~3 we present intrinsic and
projected density profiles and fit the latter structural
characteristics in terms of the $R^{1/n}$ law (\cite{ser68}). We
then proceed to illustrate in detail their phase space properties,
in particular by calculating the relevant pressure tensor
profiles, the pressure anisotropy content, the projected velocity
dispersion profiles, the line profiles (line-of-sight velocity
distribution), and the relevant phase space densities $N(E, J^2)$
and $N(E)$ (Sect.~4). In Sect.~5 we examine the stability of the
models with respect to the radial orbit instability, not only by
inspection of well-known global anisotropy indicators, but also
with the aid of a number of N-body simulations. In
Sect.~\ref{sec:fit} we perform a first test of the viability of
these models by applying them to the observed properties of
NGC 3379, a galaxy that is characterized by lack of significant
rotation and is likely to possess only small amounts of dark
matter. The conclusions are drawn in Sect.~7.

This paper also sets the basis for a detailed comparison between
the $f^{(\nu)}$ family of models and the products of collisionless
collapse resulting from N-body simulations, to be presented in a
follow-up paper. For a full comparison with the observations of
elliptical galaxies, the models should be extended to allow for
the presence of a sizable dark halo and of significant rotation.

\subsection{Models of galactic structure: physical outlook}

The need for anisotropic models to describe elliptical
galaxies can be traced back to the strong empirical arguments
that come from the observations of non-spherical geometry in the
absence of significant rotation (e.g., see the discussion provided
by \cite{ber93} and references therein); to be sure, direct
evidence for pressure anisotropy in round systems is not easy to
obtain, because so far the observed line profiles show only modest
deviations from a Gaussian (e.g., see \cite{ger01}).

The main goal of this paper is to study the properties of a given
specific dynamical context able to provide physical justification for
the existence of anisotropic equilibria, that is the dynamical
framework of collisionless collapse. The process of collisionless
collapse, together with its accompanying mechanism of incomplete
violent relaxation, is one (but not the only) element that is expected
to play an important role in the formation of stellar systems. In
spite of the many papers that have addressed issues related to such
dynamical context (in addition to the papers cited earlier in the
Introduction, e.g. see Shu 1978, 1987, Voglis 1994, Hjorth \& Madsen
1995), it is not yet clear whether an analytically tractable
distribution function, or family of distribution functions, can be
assigned to the products of such collisionless collapse. This paper
tries to provide a contribution to this problem. At the same time, we
still need to establish how far the properties of such products, often
studied in a simplified one-component picture, would be from those of
observed stellar systems, and in which way.

An enormous amount of work has focused and is currently focusing on
the demands (on galactic structure) from the cosmological context. In
particular, this has led astronomers to look for the presence, in
observed objects, of cuspy density distributions of dark matter,
following a universal profile suggested by cosmological simulations
(see Navarro, Frenk, \& White 1997; Moore et al. 1998, Ghigna et
al. 2000). Other studies have addressed the issue of the establishment
of the galaxy scaling laws (such as the Fundamental Plane for
elliptical galaxies; e.g., see Meza et al. 2003, Lanzoni et al. 2004)
in the accepted cosmological scenario. These demands are beyond the
scope of the present paper, but should eventually be faced, as a point
of contact between dynamical investigations of individual galaxies and
studies of the evolving universe from which they were formed.

In this respect, a point to be noted, to avoid unnecessary confusion
about the aims of purely dynamical studies, is the following. {\it A
priori}, studies of collisionless collapse within the line of research
adopted in this paper have nothing to say about some important issues
such as the establishment of the Fundamental Plane (but see
Gonzalez-Garcia \& van Albada 2003, Nipoti et al. 2003), because the
relevant scaling laws depend on physics that goes beyond pure
dynamics, which is inherently scale-free. In turn, pure dynamical
studies can try to explain why galaxies prefer the $R^{1/4}$ law,
which is a structural property, instead of just accepting it as an
empirical fact (as often done in a number of otherwise important
astrophysical studies).

Thus we would like to emphasize that this paper represents only one
step in the direction of a comparison with the observations. To deal
fruitfully with the presence of dark matter and other important
ingredients (such as significant rotation and the possible presence of
an additional disk component), one first has to master the properties
of one-component models, which, as shown in this paper, turn out to
exhibit a variety of interesting dynamical properties. In fact, it is
rewarding and {\it a priori unexpected} to find that, as a result of a
simple conjecture about the way to characterize incomplete violent
relaxation (the addition of Q to the natural constraints under which
Boltzmann entropy is extremized; see Sect. 2), one-component spherical
models are identified able to fit products of N-body simulations over
nine orders of magnitude in density (see Bertin \& Trenti 2004) and,
at the same time, the observed photometric profile (over about ten
magnitudes) and the inner kinematic profile (inside $R_e$) of the best
studied elliptical galaxy (see Sect.~6). Therefore, in spite of its
incompleteness, the stage reached so far is definitely interesting
from the physical point of view. Discrepancies with respect to the
observations play the welcome role of providing concrete indications
about the role of the ingredients that are {\it a priori} ignored by
the purely dynamical and highly simplified picture considered in this
paper.


\section{Model construction and the relevant parameter
space}\label{sec:model}

In the spherically symmetric limit, in order to allow for the
possibility that a stellar system is only partially relaxed, one
may extremize the Boltzmann entropy $S = - \int f \ln{f} d^3x
d^3w$ under the constraint that the total energy $E_{tot} =
(1/3)\int E f d^3x d^3w$, the total mass $M = \int f d^3x d^3w$,
and the additional quantity

\begin{equation}
Q = \int J^{\nu} |E|^{-3 \nu/4} f d^3x d^3w
\end{equation}

\noindent are constant (\cite{sti87}). Here $E = w^2/2+\Phi$ and $J^2
= |\vec{r} \times \vec{w}|^2$ represent the specific energy and the
specific angular momentum square of a single star subject to a
spherically symmetric mean potential $\Phi(r)$. (We have decided to
use the symbol $w$, instead of $v$, for the velocity variable so as to
avoid confusion with the symbol $\nu$.) If only $E_{tot}$ and $M$ were
kept fixed, the extremization process would lead to a Maxwellian, that
is to an isothermal and isotropic distribution function, appropriate
for a fully relaxed system.  Some arguments have been provided as to
why a quantity such as $Q$ should be, at least approximately,
conserved (\cite{sti87}) and those will not be repeated here. In any
case, the conservation of $Q$ should be taken as a conjecture. In a
follow-up paper (Trenti, Bertin, \& van Albada in preparation) we will
discuss this problem further by examining a set of collapse
simulations, for which the conservation of $Q$ will be tested.

Such extremization leads to the following expression for the $\f$
distribution function:

\be \label{eq:fnu}
f^{(\nu)} = A \exp {\left[- a E - d \left(
\frac{J^2}{|E|^{3/2}}\right)^{\nu/2}\right]}~,
\ee

\noindent where $\nu$, $a$, $A$, and $d$ are positive real
constants. This set of constants provides two dimensional scales
(for example a mass scale $M_{scale} = A a^{-9/4} d^{-3/\nu}$ and
a reference radius $R_{scale} = a^{-1/4} d^{-1/ \nu}$) and two
dimensionless parameters. For the latter two quantities, we may
refer to $\nu$ and $\gamma = ad^{2/\nu}/(4 \pi GA)$. The
distribution function is taken to vanish for unbound orbits, that
is for $E > 0$.

\begin{figure}
  \resizebox{\hsize}{!}{\includegraphics{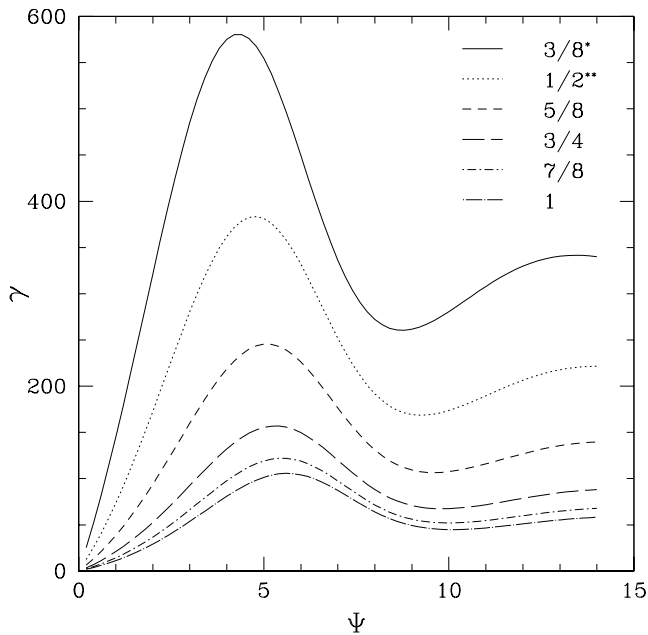}}
  \caption{Relation $\gamma(\Psi)$ for the $\f$ family of models,
  for selected values of $\nu$. To fit the adopted frame, the
  $\gamma$ values corresponding to $\nu=3/8$ have been multiplied
  by a factor $1/6$ and the ones corresponding to $\nu=1/2$ by a
  factor $2/3$.}
  \label{fig:gamma}
\end{figure}

The two-parameter family of models is then constructed by solving
the Poisson equation:

\be
\nabla^2 \Phi(\vec{r}) = 4 \pi G \int \f(\vec{r},\vec{w}) d^3\vec{w},
\ee

\noindent for the potential $\Phi (r)$. As briefly described by
Bertin \& Trenti (2003), the solution is obtained numerically
after the equation has been cast in dimensionless form:

\be
\frac{1}{\hat{r}^2} \frac{d}{d \hat{r}} \hat{r}^2 \frac{d}{d
\hat{r}} \hat{\Phi}= \frac{1}{\gamma} \hat{\rho}(\hat{r},
\hat{\Phi}),
 \ee

\noindent with $\hat{r} = r/R_{scale}$ and $\hat{\Phi} = a \Phi$
(for the adopted scaling, see the Appendix). The Poisson equation
is integrated with ``initial" conditions $d\hat{\Phi}/dr = 0$ and
$\hat{\Phi} = - \Psi$ at $r = 0$ and the parameter $\gamma$ is
determined as an eigenvalue, $\gamma = \gamma (\Psi)$, in order to
satisfy the condition of Keplerian decay ($\hat{\Phi} \sim
-1/\hat{r}$) at large radii. The general behavior of the function
$\gamma(\Psi)$ is similar to that of the corresponding function
for the $f_{\infty}$ models; after some oscillations, $\gamma$
tends to a ``plateau" at large values of $\Psi$ (as indicated by
Fig. ~\ref{fig:gamma}).

We have explored the parameter space $(\nu, \Psi)$ by means of an
equally spaced grid from $\nu=3/8$ to $\nu=1$ at steps of $1/8$,
and from $\Psi=2$ to $\Psi=13$ at steps of $0.2$. A given model
will be denoted by the values of the two parameters $(\nu;\Psi)$
in parentheses.

For given $(\nu, \Psi)$, once the solution for the potential
$\Phi(r)$ is obtained, it can be inserted in the expression of
$\f$, from which all the intrinsic and observable profiles and
properties of the model can be calculated. In particular the
density $\rho(r)$ and the anisotropy profile $\alpha(r)$, defined
as $\alpha(r) = 2 - (\langle w^2_{\theta}\rangle + \langle
w^2_{\phi}\rangle)/\langle w^2_r\rangle$, require the evaluation
of a double integral over the velocity space (see also \cite{ber03}).

The problem of constructing the line profiles $F(w,R)$ for a given
distribution function of the form $f(E,J^2)$ is often discussed in the
literature (see \cite{ger91,ger93,car95}). The calculation requires
the evaluation of a triple integral (two integrations over the
velocity space orthogonal to the line of sight and one over the radial
coordinate along the line of sight), which we have performed
numerically with the same adaptive algorithm used to compute the
density (\cite{bern91}). In practice, we have followed the procedure
described by Gerhard (1991). The calculation of the quantities
$\sigma_{proj}$ and $h_4$ is then performed as described at the end of
Sect.~\ref{sec:line}, where they are introduced and defined.

\subsection{The parameter $\Psi$ and the density concentration}

In the following we will often refer to the parameter $\Psi$ as
the {\it concentration parameter}. Strictly speaking, such term is
justified only at relatively large values of $\Psi$ ($\Psi
\ga 5$).

\begin{figure}
   \centering
   \resizebox{\hsize}{!}{\includegraphics{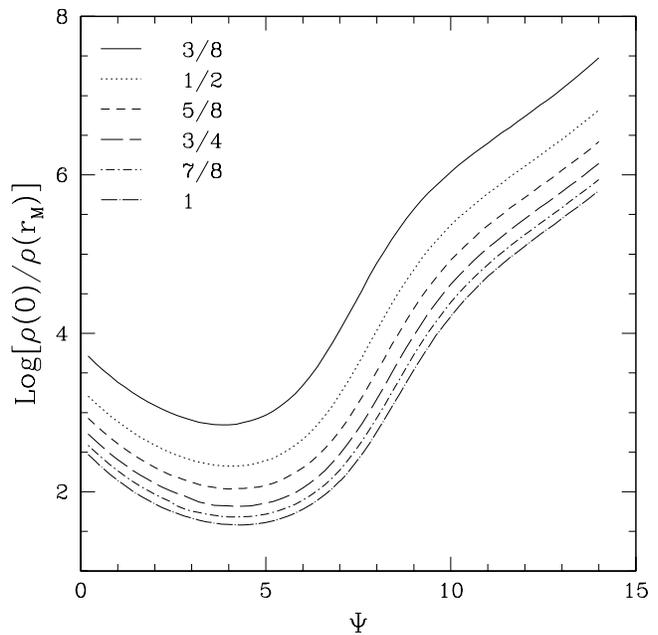}}
    \caption{Density contrast between the center and the half-mass
radius along the equilibrium sequence for different values of
$\nu$. The solid line refers to $\nu=3/8$; at fixed $\Psi$, models
with higher values of $\nu$ are less concentrated.}

         \label{fig:rho_contrast}
   \end{figure}

A more intuitive measure of the central concentration of a model
is given by the ratio $\rho(0)/\rho(r_M)$ of the central density
to the value of the density attained at the half-mass radius
$r_M$. As illustrated in Fig. ~\ref{fig:rho_contrast}, this ratio
is a monotonic increasing function of $\Psi$ only beyond a minimum
at $\Psi \approx 4.5$.

\section{Density profiles} \label{sec:dens}

In this Section we start with the discussion of the properties of
the density distribution.

\subsection{The intrinsic density profile}

\begin{figure}
  \resizebox{\hsize}{!}{\includegraphics{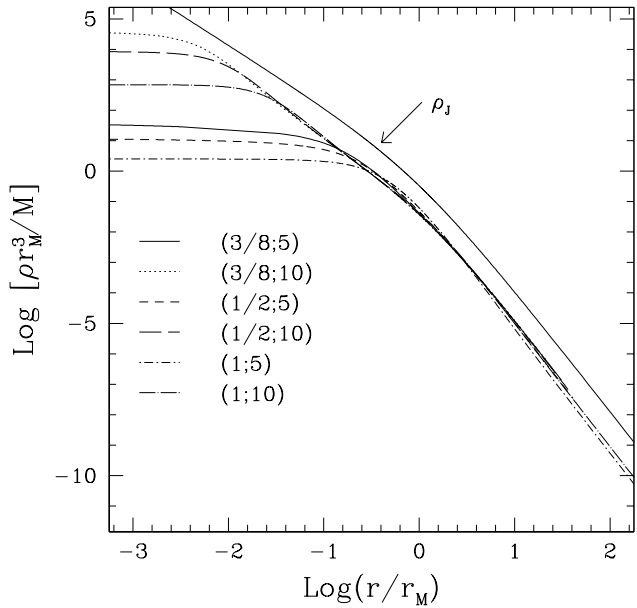}}
  \caption{Density profile of some representative $f^{(\nu)}$
  models, with $\nu = 3/8$, $1/2$, and $1$, and $\Psi = 5$
  and $10$. The most concentrated model corresponds to $(3/8;10)$
  and the least concentrated to $(1;5)$. If the scales are fixed
  so that $r_M = 10~kpc$ and $M = 10^{11}~M_{\odot}$, the units
  for the density are $10^{-1}~M_{\odot}/pc^3$. The density profiles overlap in the
  outer parts, beyond a radius that becomes smaller and smaller as
  $\Psi$ increases. In the plot we also record the $\rho_J$ profile,
  shown here on an arbitrary scale for a convenient comparison.}
  \label{fig:dens_profile}
\end{figure}

In the outer parts, i.e. at radii such that $r \ga r_M$, the
density profile (Fig.~\ref{fig:dens_profile}) is basically the
same for all models of the $\f$ family. In fact, the dimensionless
density associated with the $\f$ distribution function can be
expressed, at large values of $r$, as:

\be \label{eq:asi} \hat {\rho}(\hat{r}) = \frac{3 \pi^2}{2
\sqrt{2}} \frac{\Gamma(2/\nu)}{\nu}
\frac{1}{\hat{r}^2}\left(\hat{\Phi}^2(\hat{r})+O(\hat{\Phi}^3)
\right), \ee

\noindent where $\Gamma(x)$ is the standard Gamma function
(\cite{abr_ste}). Curiously, the $1/r^4$ behavior is found to
start from well inside the main body of the system, not too far
beyond $r_M$, while the density distribution of the inner half of
the system is similar to that of an isothermal sphere. A power-law
of the form $r^{-3.2}$ fits reasonably well the profiles in the
transition region from $r_M$ to $3r_M$.

As the concentration parameter $\Psi$ increases, the point where
the density profiles merge into a profile common to all the models
sets in at smaller and smaller radii, so that for increasing
values of $\Psi$ the models appear to converge toward a common
(singular) model with a central cusp. In these respects, the
general behavior of the intrinsic density profiles is similar to
that of the $f_{\infty}$ models. Therefore, the behavior of
concentrated models is well captured by the following simple
formula (\cite{jaf83}):

\be
\hat{\rho}_J (\hat{r}) = \frac{1}{\hat{r}^2}
\frac{1}{(1+\hat{r})^2},
\label{jaffe}
\ee

\noindent as shown in Fig.~ \ref{fig:dens_profile}. Here, and in the
remaining part of this subsection, a hat symbol denotes that a
quantity is expressed in a suitable dimensionless form, obviously with
no reference to the scaling procedure described after
Eq.~(\ref{eq:fnu}).

In turn, low-$\Psi$ models are characterized by a prominent core. In
fact, some models have density profiles close to that of the isochrone
model (\cite{hen59}) or of the so-called perfect sphere ($\hat{\rho}_p
= (1+\hat{r}^2)^{-2}$). The core may be even represented by the
density profile of a Plummer sphere (\cite{plu15}),

\be
\hat{\rho}_{Pl}(r)= \frac{1}{(1+\hat{r}^2)^{5/2}}. \ee

These density profiles are mentioned to better illustrate the
properties of the mass distribution of the $\f$ models in terms of
well-known profiles. For a discussion of the merits and limitations of
physically based models (such as the $\f$ models) with respect to
other models constructed on the basis of analytical convenience (e.g.,
see \cite{mer85}, \cite{her90}), the reader is referred to the review
article by Bertin \& Stiavelli (1993). On the other hand, in relation
to phase-space properties, we should emphasize the following point. As
will be clear from Sect.~4, in the approach adopted in this paper (in
which the distribution function is constructed from physical
arguments) the final velocity dispersion and pressure anisotropy
profiles cannot be set independently, but follow from the
self-consistent solution. In other words, all the ``observable"
profiles can be seen as consequences of the physical framework
considered. This is in sharp contrast with other modeling procedures
that are descriptive, rather than predictive (e.g., see \cite{mer85}).

\subsection{The projected density profile} \label{sec:RN}

\begin{figure}
  \resizebox{\hsize}{!}{\includegraphics{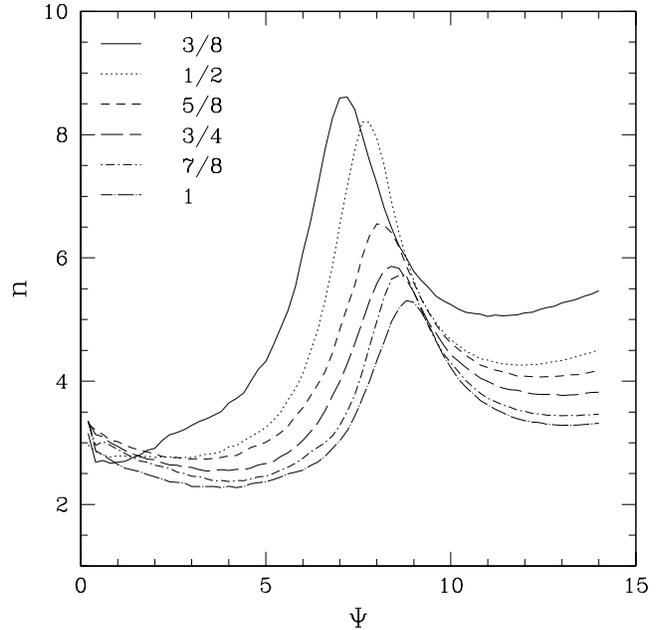}}
  \caption{Best fit value of the index $n$ (of the $R^{1/n}$ law)
  associated with the projected density profile of the $f^{(\nu)}$
  models, for selected values of $\nu$.}
  \label{fig:prolum}
\end{figure}

\begin{figure}
  \resizebox{\hsize}{!}{\includegraphics{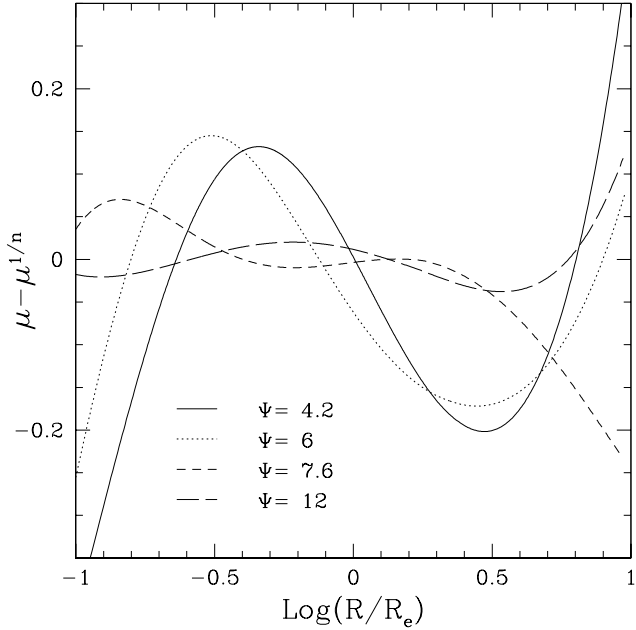}}
  \caption{Residuals $\mu -\mu^{1/n}$, in magnitudes, obtained
  by fitting the $R^{1/n}$ law to the projected density profiles of the
  $\f$ models for $\nu=1/2$ and selected values of $\Psi$.
  The quality of the fit is excellent for concentrated models ($\Psi > 7$).}
  \label{fig:resRN}
\end{figure}

We have then proceeded to compute a library of projected density
profiles (which may be compared to observed luminosity profiles,
under the assumption of a constant $M/L$ ratio). A first way to
characterize these profiles is to fit them with the $R^{1/n}$ law
(\cite{ser68}). Such fit has been performed over a very wide radial
range, from $0.1 R_{e}$ to $10 R_{e}$. It shows that the $\f$
family is well represented by the $R^{1/n}$ law, with the index
$n$ ranging from $2.5$ to $8.5$ (see Fig.~\ref{fig:prolum}; the
slightly bumpy behavior of the $\nu = 3/8$ curve just reminds us
of the uncertainties associated with the best-fit determination of
$n$). The $n = 4$ behavior, characteristic of the de Vaucouleurs
law (\cite{dev48}), is mostly associated with concentrated
high-$\Psi$ models, but we note that many intermediate-$\Psi$
models also have the same structural property. The residuals from
the $R^{1/n}$ best fit (see Fig.~\ref{fig:resRN}) are typically
within $0.05$ mag for concentrated models, while at low values of
$\Psi$ they are within $0.2$ mag; the general behavior can be
compared with that of the $f_{\infty}$ models (see Fig.~ A.1 in
\cite{ber02}).

\section{Phase space properties}

Here we describe the properties of the $\f$ models related to the
velocity space.

\subsection{Pressure anisotropy profiles and global anisotropy indicators}

\begin{figure}
  \resizebox{\hsize}{!}{\includegraphics{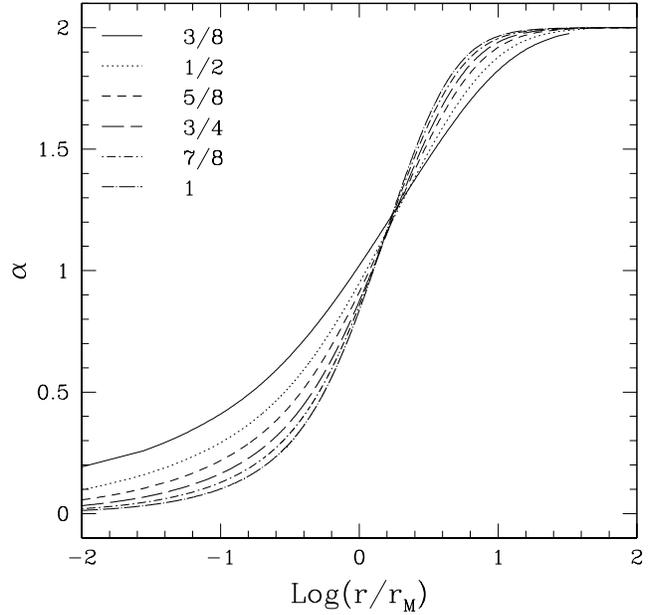}}
  \caption{Anisotropy profiles $\alpha(r)$ of $\Psi=5$ models for
  different values of $\nu$.}
  \label{fig:ani_prof}
\end{figure}

The pressure anisotropy of the models can be described by means of
the anisotropy profile $\alpha(r)$, defined as $\alpha(r) = 2 -
(\langle w^2_{\theta}\rangle + \langle w^2_{\phi}\rangle)/\langle
w^2_r\rangle$. This function, illustrated in
Fig.~\ref{fig:ani_prof}, shows that the cores are approximately
isotropic and that in the outer parts the pressure is mostly in
the radial direction, in line with the qualitative expectations of
the violent relaxation scenario (\cite{lyn67}). Higher values of
$\nu$ are associated with a sharper transition from central
isotropy to radial anisotropy.

\begin{figure}
  \resizebox{\hsize}{!}{\includegraphics{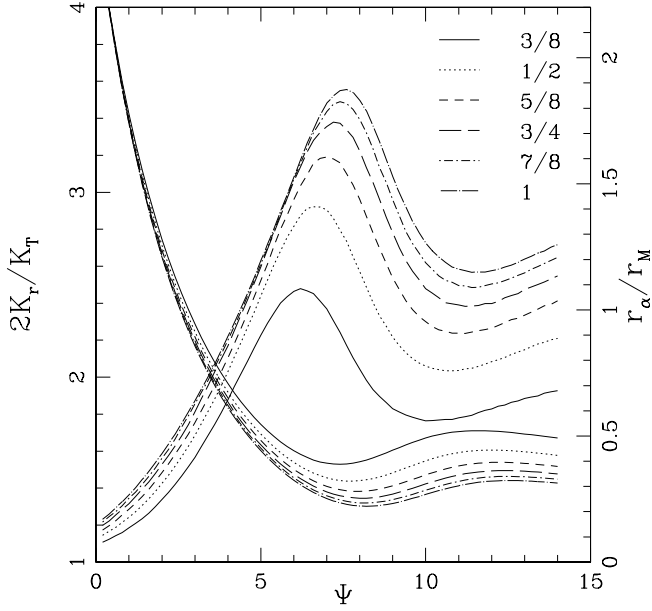}}
  \caption{Global anisotropy of the $f^{(\nu)}$ models shown in terms
  of the parameter $2K_r/K_T$ (group of curves starting from the top
  left) and of the ratio $r_{\alpha}/r_M$ (anisotropy radius to
  half-mass radius; group of curves starting from the bottom left) as
  a function of $\Psi$. For low values of $\Psi$ the global anisotropy
  $2K_r/K_T$ is basically independent of $\nu$. At fixed value of
  $\Psi$ the global anisotropy decreases with $\nu$. }
  \label{fig:ani_KrKt} \label{fig:ani_cont}
\end{figure}

The degree of anisotropy globally present in our models can be
characterized in terms of the ratio $2K_{r}/K_{T}$, where $K_r$ is
the total kinetic energy associated with the radial degree of
freedom, and $K_T$ the corresponding quantity related to the two
tangential directions. All models present an excess of kinetic
energy in the radial direction, as illustrated in
Fig.~\ref{fig:ani_KrKt}. The ratio $2K_{r}/K_{T}$ is greater than
$\approx 1.3$ over the whole sequence and becomes larger than 2 in
the low-$\Psi$ region. As a general trend, at fixed $\Psi$, $\f$
models with higher values of $\nu$ are more isotropic.

In addition to $2K_{r}/K_{T}$, as a global anisotropy indicator we
can also refer to the parameter $r_{\alpha}/r_M$ (see Fig.
~\ref{fig:ani_cont}); here the radius $r_{\alpha}$ denotes the
anisotropy radius, defined from the relation
$\alpha(r_{\alpha})=1$.  In contrast with the $f_{\infty}$ models,
for which $r_{\alpha} \approx 3 r_M$ for relatively large values
of $\Psi$, here we find that concentrated models are characterized
by $r_{\alpha} \approx r_M$.

\subsection{Velocity dispersion and projected velocity dispersion profiles}

\begin{figure}
  \resizebox{\hsize}{!}{\includegraphics{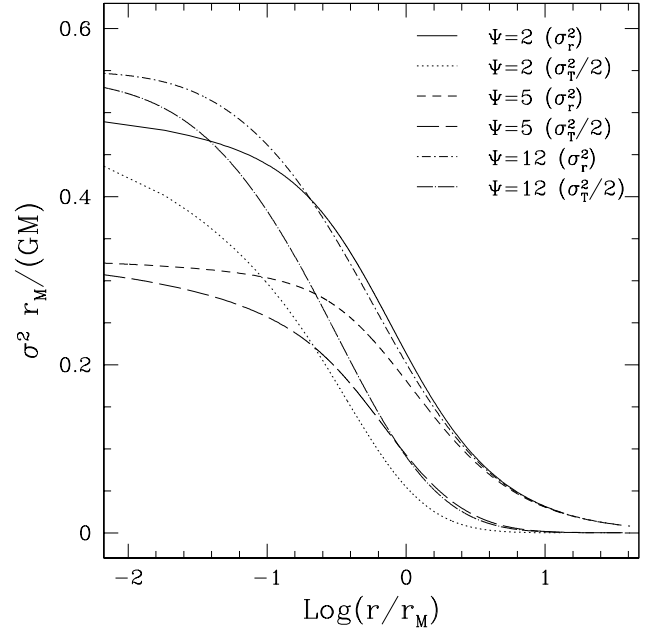}}
  \caption{Intrinsic ``pressure" profiles ($\sigma_T^2/2 = (\langle
  w^2_{\theta}\rangle + \langle w^2_{\phi}\rangle)/2$ and $\sigma_r^2
  = \langle w^2_r\rangle$) for selected $\f$ models with $\nu = 1/2$.
}  \label{fig:sigmaintr}
\end{figure}

\begin{figure}
  \resizebox{\hsize}{!}{\includegraphics{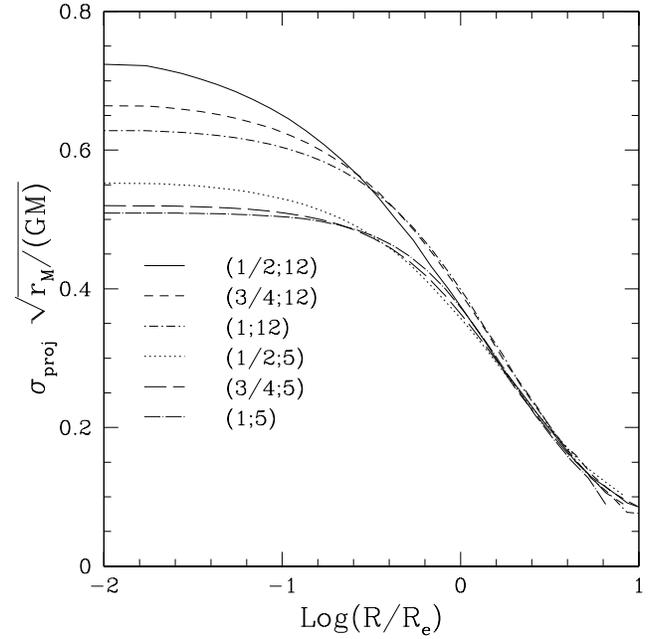}}
  \caption{Projected velocity dispersion profile for selected
  $\f$ models. If the scales are fixed so that
  $r_M = 10~kpc$ and $M = 10^{11}~M_{\odot}$, the velocity
  dispersion $\sigma_{proj}$ is given in units of $207.4~km/s$.}
  \label{fig:line_sigma}
\end{figure}

As we have seen, the models can be characterized by significant
``pressure" anisotropy. This can be illustrated directly by the
intrinsic velocity dispersion profiles, for which one may find
significant differences between the tangential $\sigma_T^2/2$ and the
radial dispersion $\sigma_r^2$ as far in as $r \approx 0.1~r_M$ (see
Fig.~\ref{fig:sigmaintr}). These velocity space properties, in
combination with the density distribution, give rise to the projected
velocity dispersion profiles (calculated from the line profiles, as
described in the next subsection), which may eventually be compared
with the observed kinematical profiles; some projected velocity
dispersion profiles are shown in Fig.~\ref{fig:line_sigma}.

\subsection{Line profiles}\label{sec:line}

Pressure anisotropy can affect the shape of the velocity
distribution integrated along a given direction, which can be
tested observationally by studying the profiles of the lines used
to determine the observed velocity dispersion. In this context,
since observational limitations prevent us from obtaining accurate
measurements of line profiles, comparisons between data and models
are often carried out in terms of certain shape parameters, which
measure the deviations from a Gaussian profile (e.g., see
\cite{ger91,SAURON}). In view of this approach, we have first
computed the line profiles for a number of models, following the
procedure described in Sect.~\ref{sec:model} and then we have
extracted from them the related value of the $h_4$ parameter
(being non rotating and spherically symmetric, the $\f$ models are
associated with line profiles characterized by vanishing $h_3$).

To extract the velocity dispersion $\sigma_{proj}$ and the value $h_4$ we
fit the line profile $F(w,R)$, at fixed $R$, with a Gaussian corrected by
a fourth order Gauss-Hermite polynomial (\cite{abr_ste}):
\begin{eqnarray}
F(w,R) &=& F_0 \exp{[-(w/\sigma_{proj})^2/2]} \times \\
& & \left\{ 1+ h_4 \left[
12-48\right(\frac{w}{\sigma_{proj}}\left)^2+ 16
\right(\frac{w}{\sigma_{proj}}\left)^4 \right ] \frac{}{}\right\}, \nonumber
\end{eqnarray}
where $F_0$, $\sigma_{proj}$, and $h_4$ are free parameters. The fit
has been performed with a Simulated Annealing method
(\cite{press}).

\begin{figure}
  \resizebox{\hsize}{!}{\includegraphics{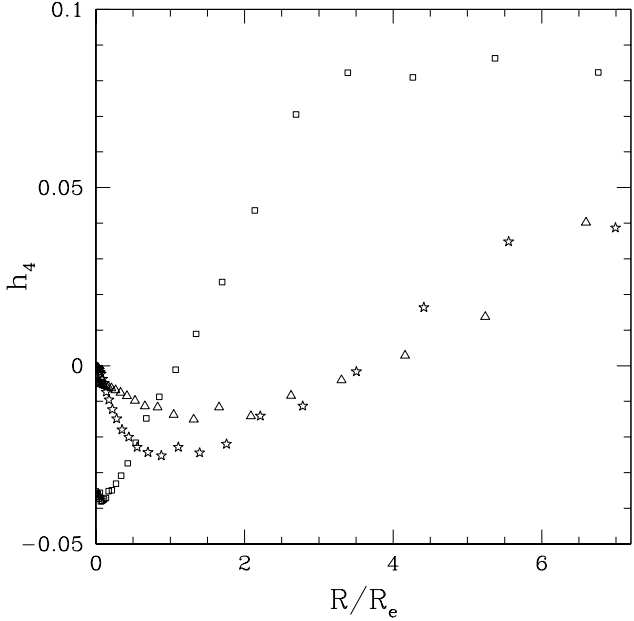}}
  \caption{Profiles of the $h_4$ parameter for selected $\f$ models
  with $\nu = 3/4$ and different values of $\Psi$ (the $\Psi = 2$
  model is denoted by squares,
  $\Psi = 6$ by triangles, and $\Psi = 10$ by stars).}
  \label{fig:h4}
\end{figure}

In general, the deviations from a Gaussian are modest, $\approx 5
\%$, and reduce to within $1 \%$ for $R \la R_{e}$. The $h_4$
parameter takes on slightly negative values at the center of the
system and then becomes positive in the outer parts (see
Fig.~\ref{fig:h4}), in line with the results found by Gerhard
(1991).

\subsection{The phase space densities $N(E, J^2)$ and $N(E)$}

\begin{figure}
  \resizebox{\hsize}{!}{\includegraphics{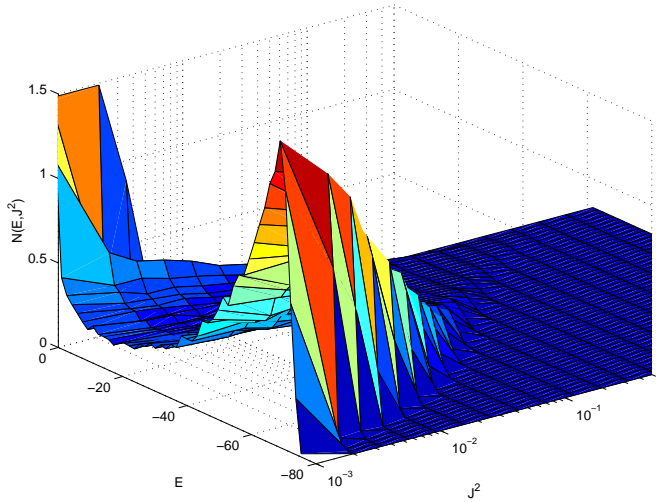}}
  \caption{The phase space density $N(E,J^2)$ for the $(1/2;6.2)$
  $\f$ model. The graph has been obtained by a Monte Carlo sampling of
  the distribution function with $2 \times 10^6$ points.}
  \label{fig:NEJ2}
\end{figure}

In view of a comparison with N-body simulations, a physically
interesting way to characterize the phase space properties of the
$\f$ models is in terms of the $N(E,J^2)$ and $N(E)$ densities,
defined in such a way that $M = \int N(E) dE = \int N(E,J^2) dE
dJ^2$. Therefore, the relation between $\f (E,J^2)$ and the phase
space density $N(E,J^2)$ is given by the Jacobian of the
transformation from $d^3\vec{x} d^3 \vec{w}$ to $dEdJ^2$:

\be
N(E,J^2)=\frac{2 \pi \f(E,J^2)}{\Omega_{r}(E,J^2)},
\ee

\noindent where $\Omega_{r}(E,J^2)$ is the radial frequency of
stellar orbits in the given potential $\Phi(r)$. At variance with
the $f_{\infty}$ models, the $\f$ models exhibit a singular
behavior of $N(E,J^2)$ near the origin in the $dEdJ^2$ phase space
(see Fig.~\ref{fig:NEJ2}).

\section{Stability} \label{sec:stability}

The study of the stability of the $\f$ models could be approached
either through a linear modal analysis (see Bertin et al. 1994) or
by means of N-body simulations. Here we prefer the latter
approach, starting from initial conditions obtained by sampling
the relevant distribution functions with Monte Carlo techniques.

\subsection{General conditions for the onset of the radial orbit instability}

A simple criterion for the onset of the radial orbit instability
states that instability occurs for systems with $2K_r/K_T \ga
1.75$ (\cite{pol81}; see also \cite{fri84}). We
thus note that models with $\Psi \la 4$ should be unstable, in a
way that is basically independent of $\nu$ (cf.
Fig.~\ref{fig:ani_KrKt}).

\subsection{Simulations of stable and unstable configurations}

We have simulated the evolution of different $\f$ models by means
of an improved version (see Trenti 2004) of a code written
originally by van Albada \& van Gorkom (1977; see also
\cite{van82}). The evolution of the system is collisionless
because the simulation particles interact with one another via a
mean field computed with Fourier techniques; the mean density
associated with the particle distribution is expanded in spherical
harmonics with contributions up to $l=4$ (the code can handle
contributions up to $l=6$). This choice of code is well suited for
our problem, since its performance has been tested to be high for
quasi-equilibrium conditions and for smooth spherically symmetric
spatial distributions. To further check our results, we have run a
comparison simulation with the tree code of Londrillo (see
\cite{lon03}) for the $(1;3.2)$ model and found no
significant differences.

In order to sample points from the $\f$ distribution function and thus
to create the initial conditions needed to study the stability of the
$\f$ models, we have used an exact inversion of the cumulative mass
profile and a two-dimensional hit/miss method in the $(w_r,
w_{\perp})$ velocity space. The angles needed to complete the
assignment of the position and velocity vectors of the simulation
particles are generated randomly. In principle, the models extend out
to infinity. In practice, we set a cut-off radius $r_{cut} \approx
50~r_M$, and we rescale the mass profile in such a way that
$M(r_{cut})= M$. The introduced mass ``loss'' is generally less than
$1\%$.

Most simulations have been run with $2 \times 10^5$ particles. If we
refer to a system with total mass $M = 10^{11}~M_{\odot}$ and
half-mass radius $r_M = 5~kpc$, the evolution is followed from $t = 0$
up to $t= t_{end} = 2~Gyr$. This corresponds to more than $20~t_d$,
with the natural dynamical time defined as $t_d=GM^{5/2}/(2K)^{3/2}$
(here $K = K_r + K_T$ is the total kinetic energy); in fact, for the
range of $(\nu;\Psi)$ considered and the above scales for $M$ and
$r_M$, we have $t_d \approx (0.65 - 0.80) \times 10^8yr$. At $t \ga 10
t_d$ the system, even when it has initially evolved because of
initial unstable conditions, settles down into an approximate
equilibrium state.

The total energy is typically conserved within $10^{-5}$ over one
dynamical time. The models are non-rotating; the total angular
momentum remains within $10^{-4}$ around the small value generated
in the initialization process.

To quantify the effects of the radial orbit instability, we have
especially focused on the evolution of the central density
concentration, of the shape of the inertia ellipsoid, computed on
the basis of the distribution of the simulation particles within a
sphere of radius $3~r_M$, and of the global anisotropy $2K_r/K_T$
ratio. The end-products of unstable initial conditions are
characterized by a final prolate state (with axial ratios $a_2/a_1
\approx a_3/a_1 < 1$), with maximum projected ellipticity
consistent with that of an $E3$ galaxy. For initially unstable
models, the central density concentration drops during evolution,
but the radial scale of the system remains approximately constant.
For example, for the $(1/2;3)$ model, the central concentration at
the end of the simulation is more than $10$ times smaller than
that of the initial conditions, but the radius of the sphere that
contains $5\%$ of the total mass increases only by $6\%$, while
the half-mass radius remains basically unchanged.

We have studied the evolution of the unstable models in terms of
exponential growth curves of the form

\be
g(t) = g_0 + g_1 [\exp{(kt)}-1],
\ee

\noindent where $g_0,~g_1$, and the growth rate $k$ are free
parameters and $g$ represents a global quantity such as the global
anisotropy ratio. To capture the initial, approximately linear
phase of the evolution, we have decided to make the fit in terms
of the exponential growth over a reduced time interval, from $t =
0$ to $t = t_{lin} < t_{end}$, defined implicitly by the relation:

\be
\left(\frac{2K_r}{K_T}\right)(t_{lin})=\frac{1}{2}\left[\left(\frac{2K_r}{K_T}\right)(t_{end})
+\left(\frac{2K_r}{K_T}\right)(0)\right]. \ee

\noindent Obviously, this choice of $t_{lin}$ is arbitrary and
should be considered only as a convenient reference time-scale.

The results of the simulations are reported in Table~1. The
threshold for instability is found to be at $2K_r/K_T \approx
1.70$, consistent with the criterion of Polyachenko \& Shukhman
(1981). Close to conditions of marginal stability, quantities such
as $2K_r/K_T$ vary slowly and by very small amounts, so that the
fit in terms of exponential growth curves is not well determined.

\begin{figure}
  \resizebox{\hsize}{!}{\includegraphics{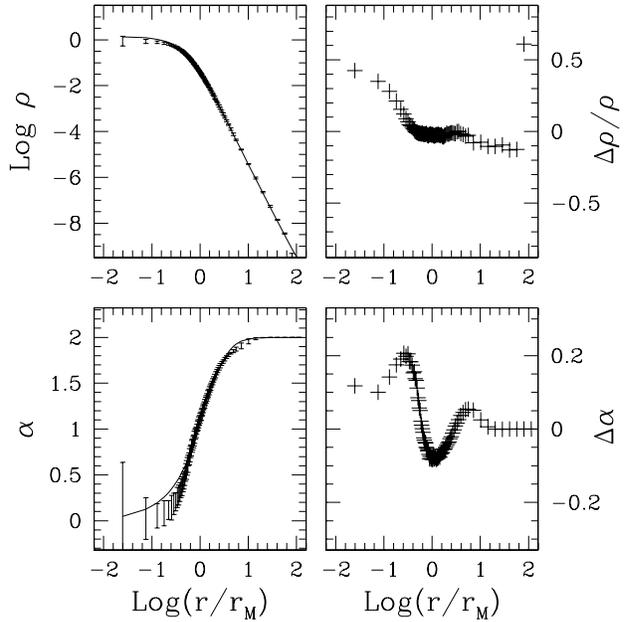}} \caption{On the
  left frames, density and anisotropy profiles of the end-products of
  a simulation starting from the unstable $(1;3.2)$ model (error bars)
  fitted by the marginally stable $(1;4.2)$ model of the $\f$ family
  (solid line); units for $\rho$ are the units adopted in the
  simulation. The right frames show the residuals $\Delta \rho =
  \rho^{(4.2)} - \rho$ and $\Delta \alpha = \alpha^{(4.2)} - \alpha$.}
  \label{fig:radorb_fit}
\end{figure}

An interesting result is the following. We checked whether the final
state reached as a result of the instability could be represented by a
model of the $\f$ family to which the initial unstable model
belonged. To do this, we fitted the spherically averaged density and
anisotropy profiles of the end-states of the simulations by means of
the same family of $\f$ models. The best fit model thus identified
(note the quality of the fit in relation to both kinematics and
density distribution) turns out to be the marginally stable model of
the sequence with the same value of $\nu$ (see also the interesting
arguments and results by \cite{pal90}). In the case of $\nu=1$,
which we studied in greatest detail, the profiles of the end-states
obtained starting from initially unstable models with $\Psi$ in the
range from 3 to 4 are well fitted by the $(1;4.2)$ model (see
Fig.~\ref{fig:radorb_fit}). For the violently unstable initial $(1;2)$
model, the final state is best reproduced by the $(7/8;4)$ model,
which is still moderately unstable.

\begin{figure}
  \resizebox{\hsize}{!}{\includegraphics{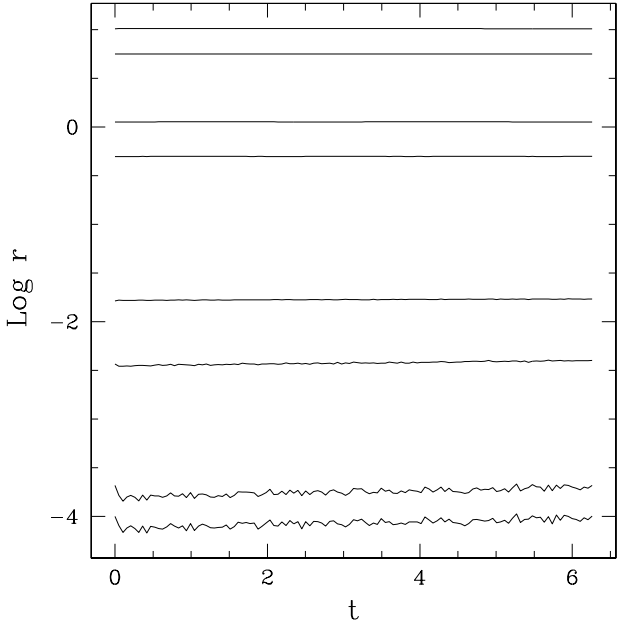}}
  \caption{Evolution of the fractional mass radii (of the sphere containing,
  from the bottom
  to the top,~$6.25 \times 10^{-5}$, $1.25 \times 10^{-4}$, $0.01$,
  $0.05$, $0.5$, $0.7$, $0.95$, $0.99$~$M$, respectively) for a simulation
  with $8 \times 10^5$ particles of the
  concentrated $\f$ model $(1/2;9.4)$. The model is stable over
  the time interval considered (about 10 dynamical times). The
  radius $r$ and the
  time $t$ are given in units of $10~kpc$ and $10^8~yr$.} \label{fig:high_conc}
\end{figure}

We have also run a number of simulations of models expected to be
stable and checked that indeed the models preserve their state for
many dynamical time scales. The most concentrated model that we
have been able to simulate properly is the $(1/2;9.4)$ model, for
which we recall that the central concentration is
$\rho(0)/\rho(r_M) = 1.14 \times 10^5$. This also quantifies the
kind of gradients and ranges of densities that our code is able to
handle (see Fig.~\ref{fig:high_conc}).

\begin{table}[!h] \label{tab:instab}
\caption{Growth rate $k$, with an estimate of the related
uncertainty $\Delta k$, for the radial orbit instability from the
simulation of a set of $\f$ models. Here $\kappa = 2K_r/K_T$ is
the global anisotropy parameter and $\eta = a_3/a_1$ is the axis
ratio of the inertia ellipsoid referred to the sphere of radius
$3~r_M$. For $M = 10^{11}~M_{\odot}$ and $r_M = 5~kpc$, $k$ is
given in units of $10^{-8}~yr^{-1}$.}
\begin{center}
\begin{tabular}{cccccc}

\hline
\hline

$(\nu;\Psi)$ &  k  & $\Delta k$& $\kappa (0) $
&$\kappa(t_{end})$&$\eta(t_{end})$  \\ \hline
(1;2.0) & 0.85 & 0.1 & 2.65 & 1.87 & 0.65 \\
(1;3.0) & 0.48 & 0.07 & 2.17 & 1.84 & 0.67 \\
(1;3.2) & 0.37 & 0.07 & 2.09 & 1.84 & 0.69 \\
(1;3.4) & 0.34 & 0.07 & 2.02 & 1.83 & 0.74 \\
(1;3.6) & 0.30 & 0.05 & 1.95 & 1.85 & 0.80 \\
(1;3.8) & 0.16 & 0.04 & 1.89 & 1.82 & 0.85 \\
(1;4.0) & 0.08 & 0.05 & 1.84 & 1.81 & 0.91 \\
(1;4.2) & 0.007 & 0.10 & 1.78 & 1.76 & 0.94 \\
(1;5.0) & 0.001 & 0.01 & 1.60 & 1.60 & 0.99\\
(1;9.0) & $< 10^{-4}$ & - & 1.32 & 1.32& 0.99 \\
(3/8;3.0) & 0.90 & 0.10 & 2.28 & 1.86 & 0.73 \\
(3/8;5.0)& 0.45 & 0.15 & 1.75 & 1.71& 0.93 \\
(1/2;3.0) &0.70  & 0.10 & 2.21 & 1.86  & 0.70 \\
(1/2;4.0) &0.43  &0.05  & 1.92 & 1.80 &0.85 \\
(1/2;5.0) & 0.1  & 0.2 & 1.68  & 1.67  & 0.96 \\
(1/2;6.0) & $< 10^{-3}$ & - & 1.54 & 1.54& 0.99 \\
(3/4;3.0) & 0.64  &0.05  & 2.20  & 1.85 & 0.70\\
(3/4;4.0) & 0.16 & 0.02 &1.86 &1.76  & 0.84  \\
(3/4;5.0) & $<2~10^{-2}$  & - &1.63  &1.63  & 0.98 \\
\hline

\end{tabular}
\end{center}
\end{table}

\section{A first comparison with the observations} \label{sec:fit}

The results in terms of the $R^{1/n}$ law described in
Sect.~\ref{sec:RN} already show that the $\f$ models possess
realistic density profiles. This encouraged us to consider a
direct comparison with an observed galaxy. For the purpose, we
picked the round elliptical galaxy \object{NGC 3379}, which apparently does
not possess significant amounts of dark matter (\cite{sag92},
\cite{rom03}; note that the models discussed in the present paper
are one-component models and thus are not applicable to systems
with prominent dark halos). This galaxy has an $R^{1/4}$
luminosity profile (\cite{dev79}, \cite{cap90}). For the
kinematics, we considered the data of Statler \& Smecker-Hane
(1999) and recently published data-points based on planetary
nebulae that extend well beyond $R_e$ (\cite{rom03}).

\begin{figure}
  \resizebox{\hsize}{!}{\includegraphics{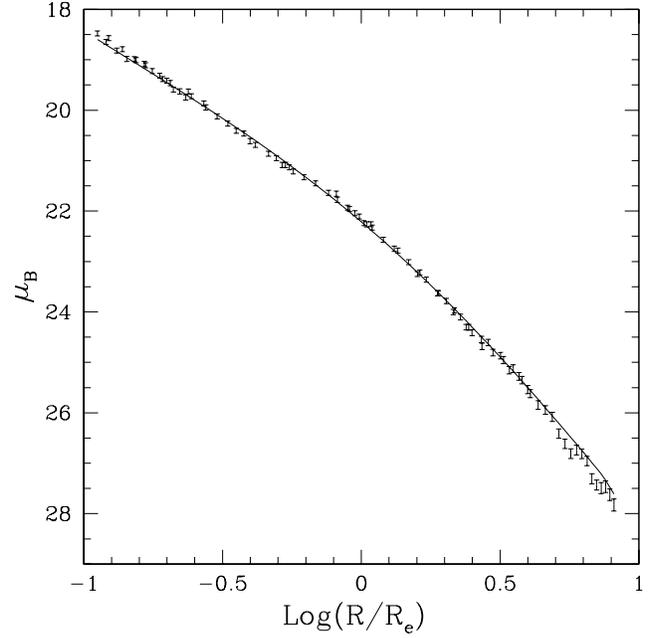}}
  \caption{Comparison between the photometric profile of \object{NGC
  3379} and the (1/2;9.4) model of the $\f$ family. The photometric
  profile in $B$ is taken from de Vaucouleurs \& Capaccioli
  (1979). Other members of the family (e.g. the (1;9.2) model) could
  better reproduce the observed data especially at large radii, but
  they would perform less well on the velocity dispersion profile.}
  \label{fig:ngc3379ph}
\end{figure}

\begin{figure}
  \resizebox{\hsize}{!}{\includegraphics{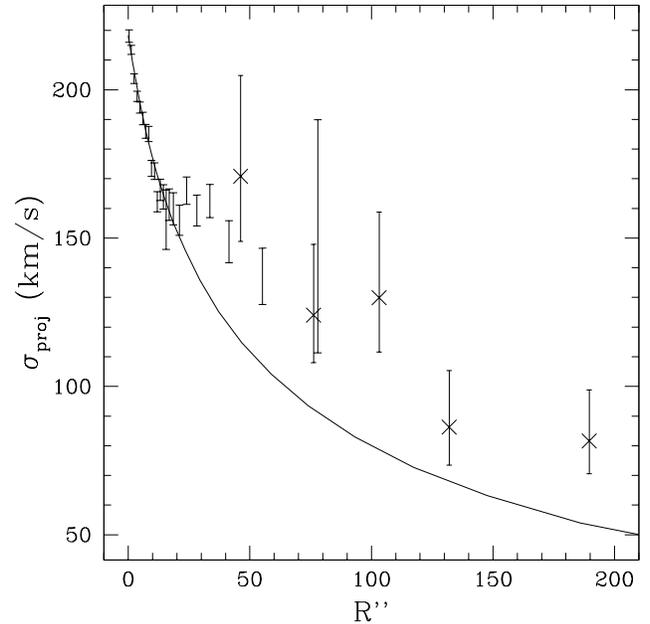}}
  \caption{Kinematic data for \object{NGC 3379} described in terms of
  the (1/2;9.4) model of the $\f$ family. The inner data (plain error
  bars) are taken from the stellar spectroscopy of Statler \&
  Smecker-Hane (1999), while the outer data (crosses with error bars)
  refer to the binned velocity dispersion determined from the study of
  planetary nebulae (\cite{rom03}). 
  }  \label{fig:ngc3379kin}
\end{figure}

The $\f$ model that best describes the data is shown in
Figs.~\ref{fig:ngc3379ph}-\ref{fig:ngc3379kin}. From such model, by
adopting a distance of $11~Mpc$ for the galaxy and an absolute
magnitude in $B$ band $M_B=-20.0$, we obtain a mass-to-light ratio
$M/L_B = 4.7$ in solar units. Population synthesis models for NGC~3379
predict a mass-to-light ratio between $4$ and $9$ (\cite{ger01}). In
comparison, Romanowsky et al. (2003) report a mass-to-light ratio
$M/L_B = 7.1 \pm 0.6$.
  
\section{Discussion and conclusions}

In this paper we have studied the structural and dynamical
properties of a two-parameter family of models of partially
relaxed stellar systems. These models had been proposed earlier
(\cite{sti87}) as physically relevant to the galaxy
formation scenario based on collisionless collapse and incomplete
violent relaxation. They had been noted to possess some realistic
properties (for $\nu$ in the range $1/2 - 1$ and for relatively
large values of $\Psi$). However, they had been left basically
aside and not studied further in systematic detail. Additional
physical interest was noted recently (\cite{ber03}), in
an investigation focused on their thermodynamic properties, in
relation to the paradigm of the gravothermal catastrophe. Because
of such physical interest, we have decided to undertake a thorough
comparison between the models and the products of collisionless
collapse, as generated in N-body simulations. Such a comparison
will be the subject of a follow-up paper, of which the present
article forms the necessary basis. In that paper we will consider
a relatively large set of simulations of collisionless collapse,
starting from a variety of initial conditions. We will address the
issue of the conservation of $Q$ and identify the range of $\nu$
for which the global quantity is best conserved. Furthermore, we
will show that in many cases the $\f$ models can provide a
surprisingly good fit to both the density and the pressure
profiles, over nine orders of magnitude of the density
distribution (see \cite{como03}). The best-fit models will
turn out to be close to marginal stability with respect to the
radial orbit instability.

Here we have shown that the family of $\f$ models exhibits a
variety of structural properties, within a common general
behavior. The model characteristics can be summarized by referring
to three separate regimes: low-$\Psi$ models (typically, $\Psi <
4$), intermediate-$\Psi$ models (typically, $4 < \Psi < 8$), and
high-$\Psi$ models (typically, $\Psi > 8$). In practice, the
values of $\Psi$ that mark the transition between different
regimes do not depend significantly on the value of $\nu$, at
least in the range explored in this paper (i.e., from $\nu = 3/8$
to $\nu = 1$).

In the intermediate-$\Psi$ and the high-$\Psi$ regimes the
concentration ratio $\rho(0)/\rho(r_M)$ increases monotonically
with $\Psi$; models with lower values of $\nu$ have larger values
of  $\rho(0)/\rho(r_M)$ (by up to one order of magnitude at given
$\Psi$, in the explored range of $\nu$). In the same regimes, as
$\Psi$ increases, the density distribution converges onto a common
profile characterized by an approximate $r^{-4}$ behavior at $r >
r_M$ and by an approximate $r^{-2}$ behavior at $r < r_M$, with an
inner core that becomes smaller and smaller. In contrast, models
in the low-$\Psi$ regime have a substantial core, comparable in
structure to that of a Plummer or of an isochrone model. In
general, the projected density distribution of the models is well
fitted by the $R^{1/n}$ law, with $n$ ranging from 2.5 to 8.5.
Typically, high-$\Psi$ models are accurately described in terms of
the $R^{1/4}$ law.

The models are all characterized by significant radially-biased
pressure anisotropy. In terms of the global anisotropy ratio
$2K_r/K_T$, at given $\Psi$ models in the low-$\Psi$ regime have
similar amounts of pressure anisotropy, which increases rapidly as
$\Psi$ decreases. In practice, all low-$\Psi$ models should all be
unstable with respect to the radial-orbit instability.
Intermediate-$\Psi$ and high-$\Psi$ models appear to be safely stable,
but they may be only marginally so for the lowest value of $\nu$ that
we considered. This latter remark also explains why we have decided
not to move below $\nu = 3/8$, given the fact that radial pressure
anisotropy is larger for lower-$\nu$ models. In terms of velocity
dispersion profiles, to some extent pressure anisotropy is already
significant at $r \approx 0.1~r_M$, even for stable models. However,
out to $r \approx r_M$ the line profiles of stable models deviate very
little from a Gaussian, with $h_4$ within the $5 \%$ level.

By means of an improved version of a code introduced earlier by
van Albada, we have checked the properties of the models with
respect to the radial-orbit instability, confirming the general
expectations based on the global anisotropy parameter $2K_r/K_T$.
Curiously, we find that unstable models tend to evolve by staying
close to the equilibrium sequence, while moving up in $\Psi$ so as
to reach marginal stability. Furthermore, we have been able to
show that the N-body code is able to handle highly concentrated
models (up to $\Psi = 9.4$, with $\rho(0)/\rho(r_M) =1.14 \times
10^5$), of which it has demonstrated the long-term stability.

Finally, a comparison with the observed surface brightness and
kinematical profiles for the galaxy NGC 3379 has shown some merits and
some limitations of the one-component family of $\f$ models when applied to
observed objects. In particular, it is interesting to see that, within
the extremely idealized framework at the basis of the construction of
these models (see also the general comments made in Sect. 1.1), the
observed luminosity profile is very well reproduced over about ten
magnitudes; at the same time the models can well reproduce the inner part
of the relevant kinematical profile, inside $R_e$. This comparison
supports the view that the models studied in this paper may be helpful to
describe the luminous component of elliptical galaxies. The obvious
limitations of the one-component models that we are considering are
clearly brought out by their inadequacy to capture the change in the
observed kinematical profile occurring around $R_e$ and thus their
failure to reproduce such profile in the outer parts. This is interpreted
as a signature of the presence of an additional component that our
models, in the form developed so far, {\it a priori} ignore.

\begin{acknowledgements}
We would like to thank T.S. van Albada and P. Londrillo, for their
collaboration and for a number of useful comments and suggestions,
and N. Douglas and A. Romanowsky, for providing us with the data
for NGC 3379. The interested reader is welcome to contact M.
Trenti, who will be glad to provide the relevant numerical
routines to initialize N-body simulations and to compute intrinsic
and projected profiles for the $\f$ family of models.
\end{acknowledgements}

\appendix

\section{Constants and scales}

From the definition of the $\f$ distribution function (see
Eq.~\ref{eq:fnu}), the constants $(A,a,d)$ have the following
dimensions:

\begin{eqnarray} \label{eq:A_dim}
& & A = [ML^{-6}T^3], \\ \label{eq:a_dim} & & a = [L^{-2}T^2],  \\
\label{eq:d_dim} & & d = [L^{-\nu /2}T^{-\nu /2}].
\end{eqnarray}

\noindent Therefore, as scales for mass and radius we may refer
to:

\begin{eqnarray} \label{eq:R_scala}
& & R_{scale}=a^{-1/4} d^{-1/ \nu}, \\ \label{eq:M_scala} & &
M_{scale}=A a^{-9/4} d^{-3/ \nu}.
\end{eqnarray}

\noindent Then we introduce the dimensionless quantities $\hat{r}
= r/R_{scale}$ (and thus $\hat{r}_M = \hat{r}_M(\nu,\Psi)=
r_M/R_{scale}$), $\hat{M}=\hat{M}(\nu,\Psi)=M/M_{scale}$, where
$\hat{M}(\nu,\Psi)=\int \hat{\rho}(\hat{r}) d^3\hat{r}$, $\hat{w}
=a^{1/2} w$, and $\hat{\Phi} = a \Phi$ for radius, mass, velocity,
and potential, respectively.

Recalling the expression for $\gamma$ (see Sect.~2, after
Eq.~(\ref{eq:fnu})), for given $\nu$ we can express $(A,a,d)$ in
terms of $(M,r_M,\Psi)$:

\begin{eqnarray}\label{eq:aa}
& & a = \frac{1}{4 \pi G \gamma(\nu,\Psi)}
\frac{r_M}{\hat{r}_M(\nu,\Psi)} \frac{\hat{M}(\nu,\Psi)}{M}, \\
\label{eq:d} & & d = [4 \pi G \gamma(\nu,\Psi)]^{\nu/4} \left(
\frac{M}{\hat{M}(\nu,\Psi)} \right)^{\nu/4} \left(
\frac{\hat{r}_M(\nu,\Psi)}{r_M}\right)^{5 \nu/4} \\ \label{eq:A} &
& A = \frac{1}{[4 \pi G \gamma(\nu,\Psi)]^{3/2}}
\left(\frac{\hat{r}_M(\nu,\Psi)}{r_M}\right)^{3/2}\left(\frac{\hat{M}(\nu,\Psi)}{M}\right)^{1/2}
\end{eqnarray}


\begin{thebibliography}{}
\expandafter\ifx\csname natexlab\endcsname\relax\def\natexlab#1{#1}\fi

\bibitem[{{Abramowitz} \& {Stegun} 1965}]{abr_ste}
{Abramowitz}, M. \& {Stegun}, I. 1965, {Handbook of Mathematical Functions}
  (Dover, New York)

\bibitem[{Antonov 1962}]{ant62}
Antonov, V.A. 1962, Vestnik Leningr. Univ., no. 19, 96 (Engl.
Transl.: in Structure and Dynamics of Elliptical Galaxies, T. de
Zeeuw, ed. Dordrecht: Reidel, 1986, 531)

\bibitem[{Berntsen {et~al.} 1991}]{bern91}
Berntsen, J., T., E., \& Genz, A. 1991, ACM Tran.Mat.Soft., 17, 437

\bibitem[{{Bertin} {et~al.} 2002}]{ber02}
{Bertin}, G., {Ciotti}, L., \& {Del Principe}, M. 2002, \aap, 386, 149

\bibitem[{{Bertin} {et~al.} 1994}]{ber94}
{Bertin}, G., {Pegoraro}, F., Rubini, F., \& Vesperini, E. 1994, \apj, 434, 94

\bibitem[{{Bertin} \& {Stiavelli} 1984}]{ber84}
{Bertin}, G. \& {Stiavelli}, M. 1984, \aap, 137, 26

\bibitem[{{Bertin} \& {Stiavelli}  1993}]{ber93}
{Bertin}, G. \& {Stiavelli}, M. 1993, Reports on Progress in
Physics, 56, 493

\bibitem[{Bertin \& Trenti 2003}]{ber03}
Bertin, G. \& Trenti, M. 2003, \apj, 584, 729

\bibitem[{Bertin \& Trenti 2004}]{como03}
Bertin, G. \& Trenti, M. 2004, AIP Conference Proc., G. Bertin, D.
Farina, R. Pozzoli eds. (American Institute of Physics; Melville,
New York), Vol. 703, pp. 318-321

\bibitem[{{Capaccioli} {et~al.} 1990}]{cap90}
{Capaccioli}, M., {Held}, E.~V., {Lorenz}, H., \& {Vietri}, M. 1990, \aj, 99,
  1813

\bibitem[{{Carollo} {et~al.} 1995}]{car95}
{Carollo}, C.~M., {de Zeeuw}, P.~T., \& {van der Marel}, R.~P. 1995, \mnras,
  276, 1131

\bibitem[{{de Vaucouleurs} 1948}]{dev48}
{de Vaucouleurs}, G. 1948, Annales d'Astrophysique, 11, 247

\bibitem[{{de Vaucouleurs} \& {Capaccioli} 1979}]{dev79}
{de Vaucouleurs}, G. \& {Capaccioli}, M. 1979, \apjs, 40, 699

\bibitem[{{de Zeeuw} {et~al.} 2002}]{SAURON}
{{de Zeeuw}, P.~T., {Bureau}, M., {Emsellem}, E., {Bacon}, R., 
	{Carollo}, C.~M., {Copin}, Y., {Davies}, R.~L., 
	{Kuntschner}, H., {Miller}, B.~W., {Monnet}, G., {Peletier},
	R.~F., \& {Verolme}, E.~K.}  2002, \mnras, 329, 513

\bibitem[Fridman \& Polyachenko 1984]{fri84}
Fridman, A.M. \& Polyachenko, V.L. 1984, Physics of Gravitating
Systems (Springer-Verlag, Berlin)

\bibitem[{Gerhard 1991}]{ger91}
Gerhard, O. 1991, \mnras, 250, 812

\bibitem[{Gerhard 1993}]{ger93}
Gerhard, O. 1993, \mnras, 265, 213

\bibitem[{{Gerhard} {et~al.} 2001}]{ger01}
{Gerhard}, O., {Kronawitter}, A., {Saglia}, R.~P., \& {Bender}, R. 2001, \aj,
  121, 1936

\bibitem[Ghigna et al.]{ghi00}
{{Ghigna}, S., {Moore}, B., {Governato}, F., {Lake}, G., 
	{Quinn}, T., \& {Stadel}, J.} 2000, \apj, 544, 616

\bibitem[Gonzalez-Garcia \& van Albada 2003]{gon03}
Gonzalez-Garcia, C. \& van Albada, T.~S. 2003, \mnras, 342, 36

\bibitem[H\'{e}non 1959]{hen59}
H\'{e}non, M. 1959, Annales d'Astrophysique, 22, 126

\bibitem[Hernquist 1990]{her90}
Hernquist, L. 1990, \apj, 356, 359

\bibitem[Hjorth \& Madsen]{hjo95}
Hjorth, J., \& Madsen, J. 1995, \apj, 445, 55

\bibitem[{{Jaffe} 1983}]{jaf83}
{Jaffe}, W. 1983, \mnras, 202, 995

\bibitem[{{Katz} 1978}]{kat78}
{Katz}, J. 1978, \mnras, 183, 765

\bibitem[1979]{kat79} Katz, J. 1979, \mnras, 189, 817

\bibitem[1980]{kat80} Katz, J. 1980, \mnras, 190, 497

\bibitem[Lanzoni et al. 2004]{lan04}  
Lanzoni, B., Ciotti, L., Cappi, A., Tormen, G., \& Zamorani, G. 2004,
\apj, 600, 640
 
\bibitem[{{Londrillo} {et~al.} 2003}]{lon03}
Londrillo, P., Nipoti, C., \& Ciotti, L. 2003, Mem. Soc. Astron.
It. Suppl., 1, 18

\bibitem[{Lynden-Bell 1967}]{lyn67}
Lynden-Bell, D. 1967, \mnras, 131, 101

\bibitem[{{Lynden-Bell} \& {Wood} 1968}]{lyn68}
{Lynden-Bell}, D. \& {Wood}, R. 1968, \mnras, 138, 495

\bibitem[Merritt 1985]{mer85}
Merritt, D. 1985, \aj, 90, 1027

\bibitem[Merritt, Tremaine, \& Johnstone 1989]{mer89}
Merritt, D., Tremaine, S., \& Johnstone, D. 1989, \mnras, 236, 829

\bibitem[Meza et al.]{mez03}
Meza, A., Navarro, J.~F., Steinmetz, M., \& Eke, V.~R. 2003, \apj, 590, 619

\bibitem[Moore et al. 1998]{moo98}
Moore, B., Governato, F. Quinn, T., Stadel, J., \& Lake, G. 1998, \apjl,
499, 5 

\bibitem[Navarro, Frenk, \& White 1997]{nfw97}
{{Navarro}, J.~F., {Frenk}, C.~S., \& {White}, S.~D.~M.} 1997, \apj,
490, 593

\bibitem[Nipoti et al. 2003]{nip03}
Nipoti, C., Londrillo, P., \& Ciotti, L. 2003, \mnras, 342, 501

\bibitem[Palmer et al. 1990]{pal90}
Palmer, P.~L., Papaloizou, J., \& Allen, A.J. 1990, \mnras, 246, 415

\bibitem[{{Plummer} 1915}]{plu15}
{Plummer}, H.~C. 1915, \mnras, 76, 107

\bibitem[{Polyachenko \& Shukhman 1981}]{pol81}
Polyachenko, V. \& Shukhman, I. 1981, Soviet Astron., 25, 533

\bibitem[{{Press} {et~al.} 1992}]{press}
Press W.~H., Teukolsky S.~A., Vetterling W.~T., \& Flannery B.~P. 1992, ``Numerical Recipes in C'', Cambridge University Press, UK 

\bibitem[{{Romanowsky} {et~al.} 2003}]{rom03} 
{{Romanowsky}, A.~J., {Douglas}, N.~G., {Arnaboldi}, M., {Kuijken},
K.,{Merrifield}, M.~R., {Napolitano}, N.~R., {Capaccioli}, M., \&
{Freeman}, K.~C.} 2003, Science, 301, 1696

\bibitem[{{Saglia} {et~al.} 1992}]{sag92}
{Saglia}, R.~P., {Bertin}, G., \& {Stiavelli}, M. 1992, \apj, 384, 433

\bibitem[Sersic 1968]{ser68} Sersic, J.L. 1968,
      Atlas de galaxias australes.
      Observatorio Astronomico, Cordoba

\bibitem[Shu 1978]{shu78}
Shu, F. 1978, \apj, 225, 83

\bibitem[Shu 1987]{shu87}
Shu, F. 1987, \apj, 316, 502

\bibitem[{{Statler} \& {Smecker-Hane} 1999}]{sta99}
{Statler}, T.~S. \& {Smecker-Hane}, T. 1999, \aj, 117, 839

\bibitem[{Stiavelli \& Bertin 1987}]{sti87}
Stiavelli, M. \& Bertin, G. 1987, \mnras, 229, 61

\bibitem[Trenti 2004]{tre04}
Trenti, M. 2004, Ph.D. Thesis, Scuola Normale Superiore, Pisa (in
progress)

\bibitem[{van Albada 1982}]{van82}
van Albada, T.~S. 1982, \mnras, 201, 939

\bibitem[van Albada \& van Gorkom 1977]{van77}
van Albada, T.,~S. \& van Gorkom, J.~H. 1977, \aap, 54, 121

\bibitem[Voglis 1994]{vog94}
Voglis, N. 1994, \mnras, 267, 379

\end{thebibliography}
\end{document}